\documentclass[article]{revtex4}

\usepackage{times}

\begin{document}

\title{The Wigner Distribution}
\author{R. F. O'Connell}
\affiliation{Department of Physics and Astronomy, Louisiana State University, Baton
Rouge, LA 70803-4001}

\maketitle

In contrast to classical physics, the language of quantum mechanics involves
operators and wave functions (or, more generally, density operators).  However, in
1932, Wigner formulated quantum mechanics in terms of a distribution function $W(q,p)$,
the marginals of which yield the correct quantum probabilities for $q$ and $p$
separately \cite{wigner}. Its usefulness stems from the fact that it provides a
re-expression of quantum mechanics in terms of classical concepts so that quantum
mechanical expectation values are now expressed as averages over phase-space
distribution functions.  In other words, statistical information is transferred from
the density operator to a quasi-classical (distribution) function.

Wigner \cite{wigner} presented a specific form for $W(p,q)$, while recognizing that
other possibilities exist, depending on the conditions which are imposed on $W$. 
Wigner's choice has the virtue of mathematical simplicity but it has the feature
that it may take negative values, with the result that several authors have
investigated non-negative distribution functions.  However, we regard negative values
of $W$ as a manifestation of its quantum nature and the fact that it " - - - cannot
be really interpreted as the simultaneous probability for coordinates and momenta - -
-".  \cite{wigner} Wigner's original paper was concerned with using $W$ for the
specific purpose of calculating the quantum correction for thermodynamic
equilibrium.  The recognition of its more general applicability stems mainly from the
work of Groenewold \cite{groen} and Moyal \cite{moyal}, who investigated the
correspondence between physical quantities and quantum operators and showed, in
particular, that the correspondence is not unique and moreover, that the distribution
functions obtained by the Weyl correspondence \cite{weyl} are the Wigner functions. 
Moyal also showed how the time dependence of $W$ and other such functions (which
arise from alternative association rules other than Wigner-Weyl but which lead to the
same physical results)  may be determined without using the Schr$\ddot{o}$dinger
equation.  In fact, Moyal's paper was a landmark contribution as, in essence, " - -
it establishes an independent formulation of quantum mechanics in phase space".
\cite{zachos}  As for all quantum formulations, Ballentine \cite{ballentine} has shown
that the development of the classical limit of the Wigner distribution is a subtle
process, especially in view of the fact that, in general, $W(q, p)$ has negative parts. 
Turning to specifics, we present some basic results developed in the original pioneering
papers \cite{wigner, groen, moyal, weyl} but conveniently presented in a comprehensive
review by Hillery et al. \cite{hill}.  Thus, in one dimensional space (generalization to
$n$ dimensions being straightforward), for a mixed state represented by a densty matrix
$\hat{\rho}$,

\begin{equation}
W(q,
p)=\frac{1}{\pi\hbar}\int^{\infty}_{-\infty}~dy\langle{q}-y|\hat{\rho}|q+y\rangle
{e}^{2ipy/\hbar}, \label{wd1}
\end{equation} whereas, for a pure state represented by a wave function $\psi(q)$,

\begin{equation}
W(q,
p)=\frac{1}{\pi\hbar}\int^{\infty}_{-\infty}~dy\psi^{\ast}(q+y)\psi(q-y){e}^{2ipy/\hbar}.
\label{wd2}
\end{equation}  However, in order to calculate correct expectation values and ensemble
averages, it is also necessay to specify the classical function $A(q, p)$ corresponding
to a quantum operator $\hat{A}$ as

\begin{equation}
A(q, p)=\int~dz~e^{ipz/\hbar}\langle{q}-\frac{1}{2}z|\hat{A}|q+\frac{1}{2}z\rangle,
\label{wd3}
\end{equation} so that $\int\int~dq~dp~A(q, p)=2\pi\hbar~~Tr(\hat{A})$.  This ensures
that

\begin{equation}
\int~dq~\int~dp~A(q, p)B(q, p)=(2\pi\hbar)~Tr(\hat{A}\hat{B}), \label{wd4}
\end{equation} and

\begin{equation}
\int~dq~\int~dp~A(q, p)W(q, p)=Tr(\hat{\rho}\hat{A}(\hat{q}, \hat{p})), \label{wd5}
\end{equation} so that, in particular, we see that $W(q, p)$ derived from the density
matrix, is $(2\pi\hbar)^{-1}$ times the phase space operator which corresponds to the
same matrix.

Following these original papers, \cite{wigner,groen,moyal,weyl} there were
many papers devoted to extending the framework and overall understanding of
distribution functions.  In addition, distributions other than those of Wigner were
introduced, notable those of Kirkwood, Cahill and Glauber, Glauber, Sudarshan and
Husimi (all of which are reviewed in Ref. \cite{oconnell}, where it is noted that some
of these are everywhere non-negative) and Cohen \cite{cohen66} and all require classical
functions different from that given in (\ref{wd3}) in order to ensure consistency.  It
is clear that all distribution functions are not measurable, despite some claims to the
contrary in the literature, where in fact what is observed are the marginal $q$
probabilities from which values of $W(q, p)$ are inferred but one could equally have
inferred values for other distribution functions.

The earliest applications of the Wigner function were in the arena of statistical
mechanics but, more recently, among the diverse areas in which the
$W$ function was found to be useful we mention hydrodynamics \cite{irving}, plasmas
\cite{brittin}, quantum corrections for transport coefficients \cite{choi}, collision
theory \cite{carr} and signal analysis
\cite{cohen95}.  However, we feel that the overwhelming majority of applications are to
be found in quantum systems where fluctuations and dissipation are playing an important
role.  In this context, the 1984 review of the $W$ function by Hillery et
al.
\cite{hill} made extensive reference to its relevance in quantum optics, which is
underlined by the more recent books of Scully and Zubairy \cite{scully} and Schleich
\cite{schleich}.  Complementary to this work is the application of the $W$ function
to a variety of problems in quantum statistical mechanics, where effects associated
with the analysis of quantum systems in a heat bath (including the radiation field
heat bath) are of the essence.  As examples of the usefulness of the $W$ function in
this context we note its role in obtaining the simplest approach to solving the
initial value quantum Langevin equation and, concomitantly, the solution to an exact
master equation \cite{ford} and also its role in the investigation of
Schr$\ddot{o}$dinger cat superpositions \cite{ford2}.  However there are
limitations to the usefulness of the $W$ function (some of which were discussed
by Moyal \cite{moyal}), notably for particles with spin and for relativistic
particles.  Finally, we mention the excellent and comprehensive overview of selected
papers on quantum mechanics in phase space, with emphasis on the Wigner function
\cite{zachos}.

\end{document}